\documentclass[pra,aps,twocolumn]{revtex4-1}
\usepackage{amsmath}
\usepackage{amssymb}
\usepackage{graphicx}

\newcommand{\mP}{{\mathcal P}}
\newcommand{\mT}{{\mathcal T}}

\newcommand{\beqa}{\begin{eqnarray}}
\newcommand{\eeqa}{\end{eqnarray}}

\renewcommand{\Im}{{\rm Im}}

\begin{document}
\title{$\mP\mT$-symmetry breaking and universal chirality in a $\mP\mT$-symmetric ring} 
\author{Derek D. Scott}
\author{Yogesh N. Joglekar}
\affiliation{Department of Physics, 
Indiana University Purdue University Indianapolis (IUPUI), 
Indianapolis, Indiana 46202, USA}
\date{\today}
\begin{abstract} 
We investigate the properties of an $N$-site tight-binding lattice with periodic boundary condition (PBC) in the presence of a pair of gain and loss impurities $\pm i\gamma$, and two tunneling amplitudes $t_0,t_b$ that are constant along the two paths that connect them. We show that the parity and time-reversal ($\mP\mT$)-symmetric phase of the lattice with PBC is robust, insensitive to the distance between the impurities, and that the critical impurity strength for $\mP\mT$-symmetry breaking is given by $\gamma_{PT}=|t_0-t_b|$. We study the time-evolution of a typical wave packet, initially localized on a single site, across the $\mP\mT$-symmetric phase boundary. We find that it acquires chirality with increasing $\gamma$, 
and the chirality reaches a universal maximum value at the threshold, $\gamma=\gamma_{PT}$, irrespective of the initial location of the wave packet or the lattice parameters. Our results imply 
that $\mP\mT$-symmetry breaking on a lattice with PBC has consequences that have no counterpart in open chains. 
\end{abstract}
\maketitle

{\it Introduction.} Recently, lattice models with a non-Hermitian Hamiltonian that is parity and time-reversal ($\mP\mT$)-symmetric, and their experimental realizations in optical waveguides~\cite{expt1,expt2} and coupled resistor-inductor-capacitor (RLC) circuits with balanced loss and gain~\cite{expt3}, have become a focal point of research. Properties of continuum, non-Hermitian, $\mP\mT$-symmetric Hamiltonians were first investigated by Bender and co-workers more than a decade ago~\cite{bender1,bender2}; they showed that although such a Hamiltonian is not Hermitian, it has purely real eigenvalues over a range of parameters. Since then, significant theoretical progress on non-Hermitian, $\mP\mT$-symmetric Hamiltonians in continuum theories and lattice models has occurred~\cite{bender3,mostafa,znojil,bendix,spin}. Traditionally, the region of parameters where the spectrum of the Hamiltonian is real and its eigenfunctions are simultaneous eigenfunctions of the combined $\mP\mT$-operation is defined as the $\mP\mT$-symmetric region, and the emergence of first pair of complex (conjugate) eigenvalues is defined as $\mP\mT$ symmetry breaking. During the past two years, one-dimensional, tight-binding lattice models with $\mP\mT$-symmetric, position-dependent tunneling amplitudes and on-site potentials have been extensively investigated~\cite{song0,song1,longhi,mark}. These investigations have shown that although the $\mP\mT$-symmetric phase is exponentially fragile in a constant-tunneling-model~\cite{bendix}, it can be strengthened by choosing appropriate tunneling profiles~\cite{localpt} and that the degree of $\mP\mT$-symmetry breaking in such systems can be tuned by the tunneling profile~\cite{derek}. The equivalence between a $\mP\mT$-symmetric, non-Hermitian system and a Hermitian system, too, has been demonstrated in limited cases~\cite{song1,song2,ya}. 

In the past decade, evanescently coupled optical waveguides~\cite{christo} have provided a unique realization of a one dimensional lattice with tunable tunneling amplitudes~\cite{gf}, disorder~\cite{berg1}, and non-Hermitian, on-site, impurity potentials~\cite{makris}. Today, these systems are a promising candidate for experimental investigations of $\mP\mT$-symmetry breaking phenomena in lattice models~\cite{expt1,expt2}. In these systems, the tunneling is controlled by the width of the barrier between adjacent waveguides and the complex, on-site potential is determined by the local refractive index. The number of waveguides in a typical array is $N\sim 10-100$, and the experimental realizations correspond to a lattice with open boundary condition. Therefore, most theoretical investigations have focused on truncated open lattices, but properties of chains with periodic boundary conditions (PBC) have not been explored; for a notable exception, see~\cite{znojilring}. We remind the reader that although the differences between properties of open chains and lattices with PBC vanish as $N\rightarrow\infty$, for small $N\lesssim 100$, relevant here, they can be significant. 

In this paper, we investigate an $N$-site lattice with PBC in the presence of a pair of non-Hermitian impurities $\pm i\gamma$. Note that on a lattice with PBC, such a pair represents $\mP\mT$-symmetric impurities irrespective of the individual impurity locations; in other words, for given impurity locations, a ``parity" operator can be defined such that the 
impurity Hamiltonian becomes $\mP\mT$-symmetric. The lattice is characterized by two tunneling amplitudes that are uniform along the two paths that connect the impurities, but may be different from each other. 

Our salient results are as follows: i) The $\mP\mT$-symmetric phase is robust~\cite{caveat} and insensitive to the distance between the non-Hermitian impurities. ii)  The critical impurity strength $\gamma_{PT}$ at which the $\mP\mT$-symmetry breaks is equal to the difference between the tunneling amplitudes along the two paths, and is thus widely tunable. iii) As $\gamma$ increases from zero, a time-evolved wave packet acquires chirality that is independent of the initial wave packet location or the distance between the impurities. iv) This chirality, quantified by a steady-state, dimensionless momentum $p(\gamma)$ reaches a universal maximum at the $\mP\mT$-breaking point, $p(\gamma_{PT})=1$. For $\gamma\geq\gamma_{PT}$, although the norm of the wave function, or equivalently, the net intensity increases exponentially with time~\cite{kottospower}, the chirality is reduced. Our results predict that $\mP\mT$-symmetry breaking in a lattice system with PBC is accompanied by novel signatures that have no counterpart in open chains. 

{\it Tight-binding model.} We start with a one-dimensional chain with $N$ lattice sites and periodic boundary conditions. Without loss of generality, we take the gain and loss impurities $(i\gamma,-i\gamma)$ at positions $(1,d)$ where $2\leq d\leq N$. The Hamiltonian for this chain is given by $\hat{H}_{PT}=\hat{H}_0 +\hat{V}$ where the Hermitian tunneling Hamiltonian $\hat{H}_0$ is given by 
\begin{eqnarray}
\label{eq:h0}
\hat{H}_0 & = & -\sum_{i=1}^{N} t(i)\left(a^{\dagger}_{i+1} a_i + a^{\dagger}_i a_{i+1}\right),\\
t(i) & = & \left\{
\begin{array}{cc}
t_b >0 & 1\leq i < d,\\
t_0 >0 & d\leq i\leq N.\\
\end{array}\right.
\end{eqnarray}
Here $t(j)$ is the tunneling amplitude between adjacent sites $j$ and $j+1$, $a^\dagger_j$ ($a_j$) is the creation (annihilation) operator for a single-particle state $|j\rangle$ localized at site $j$, and periodic boundary conditions imply that $a^\dagger_{N+1}=a^\dagger_1$. The $\mP\mT$-symmetric, non-Hermitian potential is given by 
\begin{equation}
\label{eq:v}
\hat{V}=i\gamma\left(a^\dagger_1a_1-a^{\dagger}_d a_d\right)\neq \hat{V}^\dagger.
\end{equation}

When $\gamma=0$, the energy spectrum of the Hamiltonian $\hat{H}_{PT}$ is given by $E(k,k')=-2t_0\cos(k)=-2t_b\cos(k')$, and is bounded by $2\max(t_0,t_b)$. When $t_0\geq t_b$, the $N$ eigenmomenta correspond to purely real $k$ values and $k'$ values that are either real or purely imaginary; when $t_0\leq t_b$, the situation is reversed~\cite{jake}. When $\gamma\neq 0$, since the tunneling amplitudes along the two paths between the source and sink are constant, an arbitrary eigenfunction $|\psi\rangle=\sum_j f_j|j\rangle$ with energy $E(k,k')$ can be expressed using the Bethe ansatz as~\cite{song0,jake} 
\begin{equation}
f_n=\left\{
\begin{array}{cc}
A\sin(k'n)+B\cos(k'n) & 1\leq n\leq d,\\
P\sin(kn)+Q\cos(kn) & d+1\leq n\leq N.\\
\end{array}
\right.
\end{equation}
By considering the eigenvalue equation $\hat{H}_{PT}|\psi\rangle=E|\psi\rangle$ near impurity locations, we find that the dimensionless quasimomentum $k$ (or equivalently $k'$) obeys the following equation~\cite{jake}
\begin{eqnarray}
M(k,k') & \equiv & t_0^2\sin[k'(d-1)]\sin[k(N-d-1)]\nonumber\\
& +&  t_b^2\sin[k'(d+1)]\sin[k(N-d+1)]\nonumber\\
& - & 2t_bt_0\left(\sin(k'd)\sin[k(N-d)]+\sin(k')\sin(k)\right)\nonumber\\
\label{eq:m}
& + &\gamma^2\sin[k'(d-1)]\sin[k(N-d+1)]=0.
\end{eqnarray}
When $t_0\geq t_b$, the real quasimomenta $k$ are constrained by $-\pi<k\leq\pi$, and states with quasimomenta $k$ and $-k$ are degenerate except when $k=0,\pi$, provided the corresponding $k'$ is purely real. If $k_0$ is a real quasimomentum, then $\pi-k_0$ is also a quasimomentum if and only if $N$ is even; when $t_0\leq t_b$, the situation is reversed. Thus, in contrast with an open chain, the spectrum of a chain with PBC has a particle-hole symmetric spectrum if and only if $N$ is even. 

\begin{figure}[t!]
\begin{center}
\includegraphics[angle=0,width=\columnwidth]{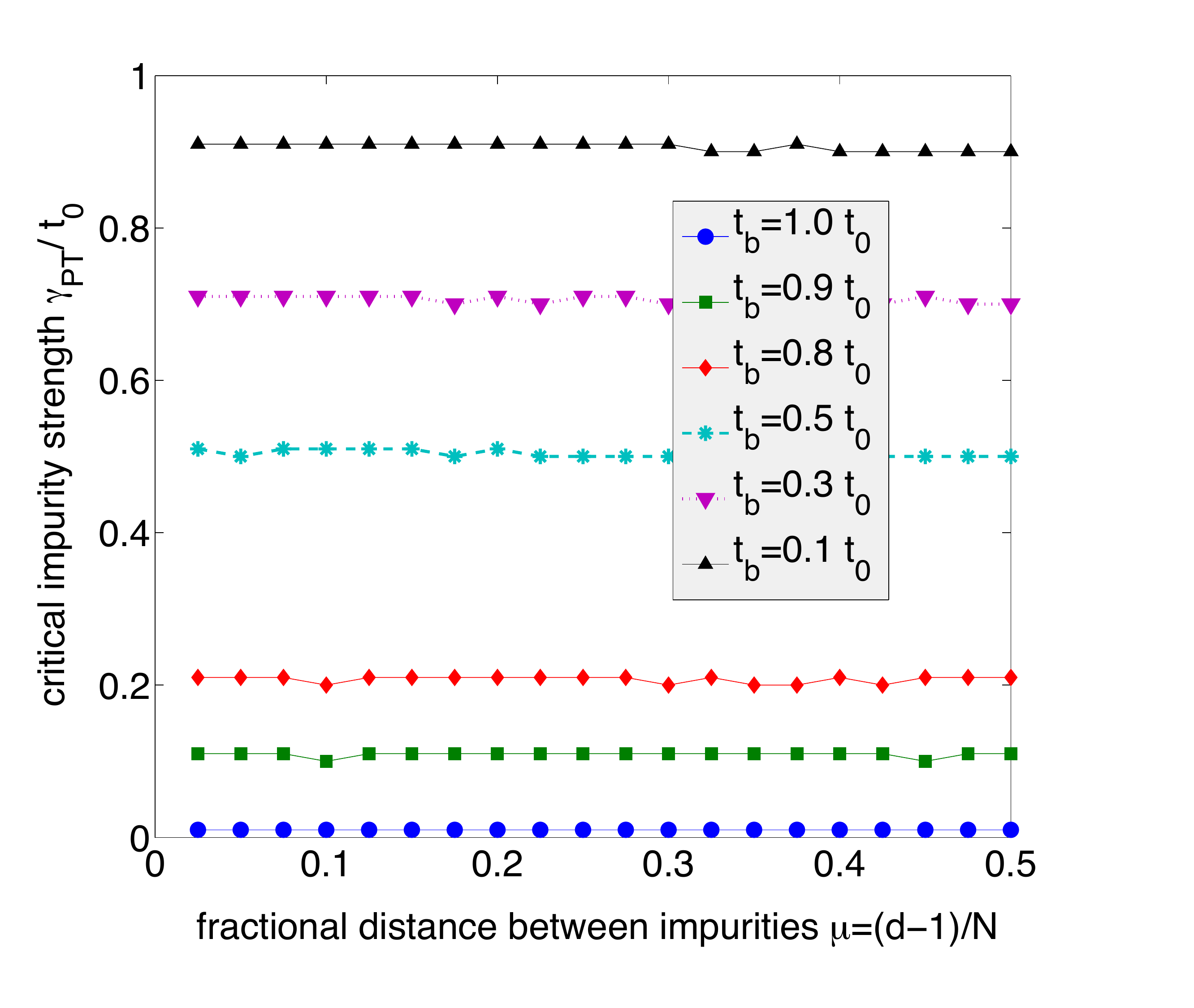}
\caption{(color online) Typical $\mP\mT$ phase diagram of a lattice with PBC. These results are for a chain with $N=40, t_0=1$ and impurities $\pm i\gamma$ at sites $(1,d)$ for different values of the sink-position $d$ and different $t_b\leq t_0$; we get identical results for odd $N$ or $t_b> t_b$. Remarkably, the critical impurity strength $\gamma_{PT}(\mu)=|t_0-t_b|$ is independent of the fractional distance $\mu$ between the impurities. Thus, the $\mP\mT$-symmetric region in a chain with PBC is more tunable that its open chain counterpart.}
\label{fig:ptphase}
\end{center}
\vspace{-5mm}
\end{figure}
Figure~\ref{fig:ptphase} shows the numerically obtained typical phase diagram for the critical impurity strength $\gamma_{PT}$ as a function of fractional distance between impurities $\mu=(d-1)/N$. Note that since there are two paths from the source $i\gamma$ to the sink $-i\gamma$, we restrict the fractional distance to $0<\mu\leq 1/2$. These results are obtained for a chain with $N=40, t_0=1$ and $t_b\leq t_0$. We obtain identical results for odd $N$ and different values of tunneling amplitudes including $t_b>t_0$, when the impurity strength $\gamma$ is measured in the units of $\max(t_0,t_b)$. We also find that the weak $\mu$-dependence of the critical impurity strength $\gamma_{PT}/\max(t_0,t_b)$ vanishes as $N$ increases.

This remarkable phase diagram predicts that in a chain with PBC, {\it the critical impurity strength $\gamma_{PT}(\mu)$ is independent of the inter-impurity distance $\mu$, and is given by} $\gamma_{PT}=|t_b-t_0|$.  It implies that the fragile $\mP\mT$-symmetric phase in an open chain with constant tunneling~\cite{bendix,mark} is stabilized and strengthened by periodic boundary condition, and that the critical impurity strength $\gamma_{PT}$ can be easily tuned by an appropriate choice of tunneling amplitudes. 

To gain insight into the insensitivity of $\gamma_{PT}(\mu)$ to the inter-impurity distance $\mu$ for large $N\gg 1$, let us consider Eq.(\ref{eq:m}) in the limit $1\ll d\ll N$, 
\begin{equation}
\label{eq:newm}
\frac{\sin(k'\mu N)\sin\left[k(1-\mu)N\right]}{\sin(k')\sin(k)}=\frac{2t_0t_b}{\left[(t_0-t_b)^2+\gamma^2\right]}.
\end{equation}
Graphical solutions of Eq.(\ref{eq:newm}) show that as $\gamma$ increases, irrespective of $\mu$, two adjacent quasimomenta near $k\sim \pi/2$ become degenerate and then complex, leading to the $\mP\mT$-symmetry breaking. 

\begin{figure*}[htb]
\begin{center}
\begin{minipage}{20cm}
\begin{minipage}{9cm}
\hspace{-2cm}
\includegraphics[angle=0,width=9cm]{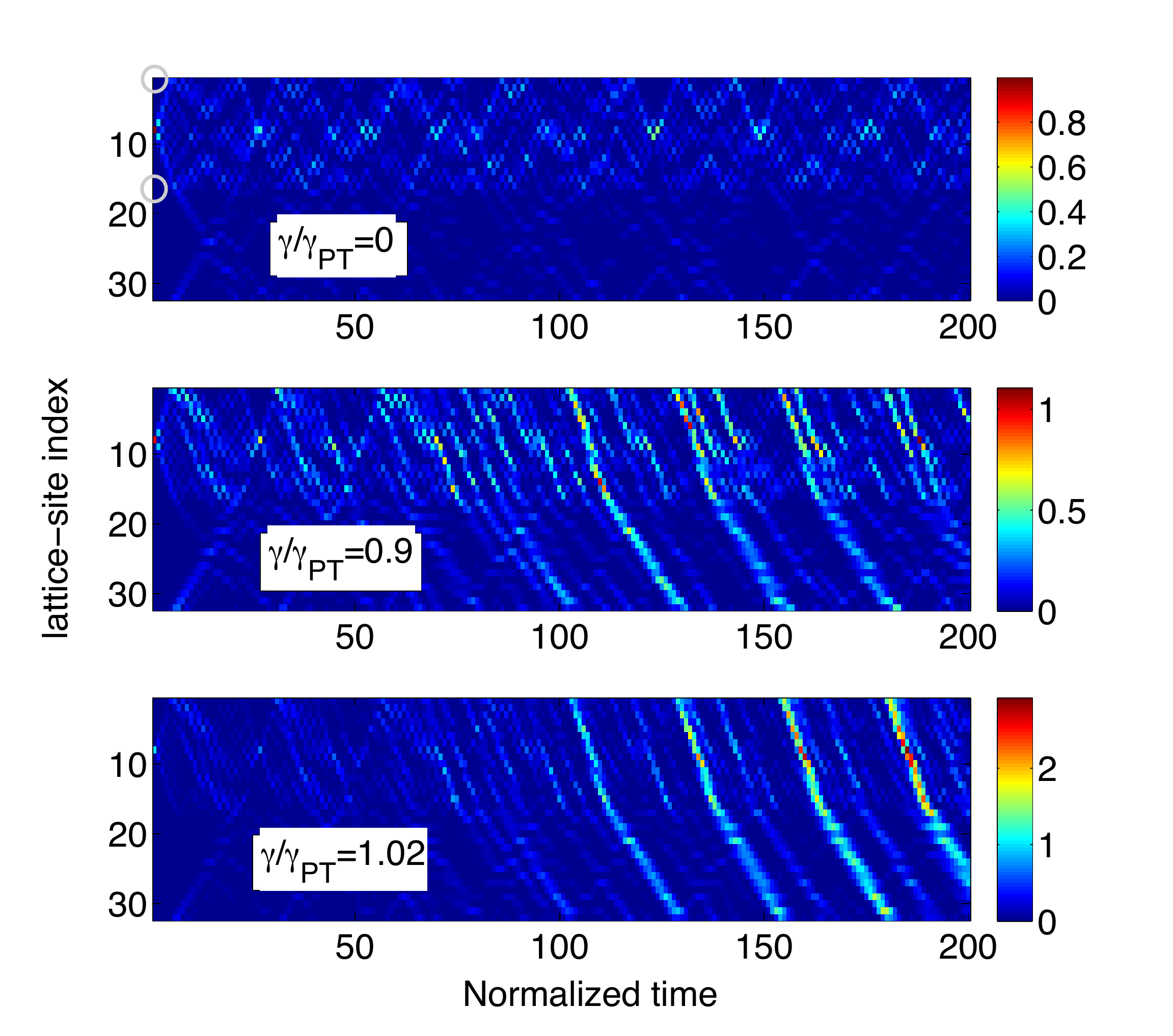}
\end{minipage}
\begin{minipage}{9cm}
\hspace{-2cm}
\includegraphics[angle=0,width=9cm]{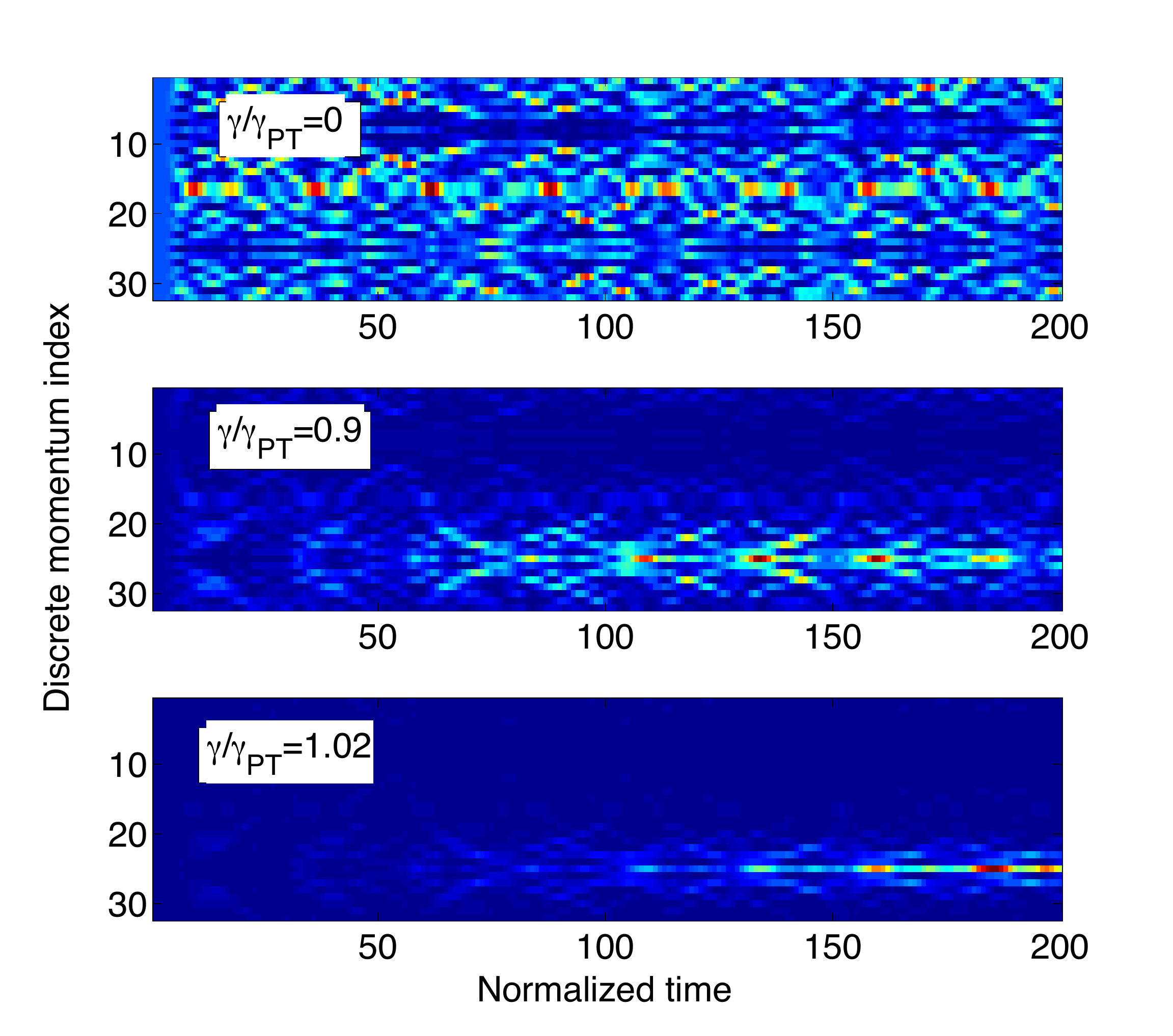}
\end{minipage}
\end{minipage}
\caption{(color online) The left-hand column shows the evolution of real-space intensity $I_R(j,t)$ with increasing $\gamma\geq 0$ in an $N=32$ lattice with impurities $\pm i\gamma$ at sites 1 and 16, shown by gray circles in the top panel, $t_0=0.5$, $t_b=1.0$, and initial wave packet localized at site $m_0=8$. When $\gamma=0$ the wave packet diffuses without preferential chirality. As $\gamma$ increases towards $\gamma_{PT}$ (center panel) and beyond (bottom panel), the wave function evolution becomes chiral. Recall that due to PBC, the bottom-most lattice site in each panel is connected with the top-most lattice site. The right-hand column shows corresponding reciprocal-space intensity $I_M(u,t)$. The vertical axis, $1\leq u\leq N$ corresponds to reciprocal space-index $-\pi< p_u\leq\pi$. As $\gamma$ increases, we see that the reciprocal-space intensity develops a clear peak at a finite, positive value of $p_u$, consistent with the chiral motion of the wave packet in the left-hand column.}   
\label{fig:psi}
\end{center}
\vspace{-5mm}
\end{figure*} 

{\it Wave function dynamics.} Now we consider the real- and reciprocal-space time evolution of a wave function that is initially localized on a single site. In an optical-waveguide realization of a $\mP\mT$-symmetric system, this initial state is most easily achievable. For an arbitrary, normalized initial state $|\psi(0)\rangle$, the wave function at time $t$ is given by $|\psi(t)\rangle=\exp(-i \hat{H}_{PT}t/\hbar)|\psi(0)\rangle$ where $\hbar$ is the scaled Planck's constant and the time evolution operator $\exp(-i\hat{H}_{PT}t/\hbar)$ is not unitary since the Hamiltonian $\hat{H}_{PT}$  is not Hermitian. We denote the site- and time-dependent real-space intensity by $I_R(j,t)=|\langle j|\psi(t)\rangle|^2$ where $j=1,\ldots, N$ denotes the site index, and use $I_M(u,t)=|\langle u |\psi(t)\rangle|^2$ to denote the reciprocal-space intensity where the discrete index $u=1,\ldots,N$ corresponds to reciprocal-space index $p_u=\pi(2u/N-1)$ with $-\pi<p_u\leq\pi$. 

The left-hand column in Fig.~\ref{fig:psi} shows the typical evolution of real-space intensity with increasing impurity strength. These results are for a lattice with $N=32, t_0=0.5, t_b=1.0$, the source and sink at sites 1 and $d=16$ respectively, and the initial wave packet localized on site $m_0=8$. The vertical axis in each panel indicates the site index, and the horizontal axis denotes time measured in units of $2\pi\hbar/\max(t_0,t_b)$. When $\gamma=0$ (top panel), the wave packet diffuses, suffering a change in speed at the impurity locations consistent with the change in the tunneling amplitude.  As $\gamma$ increases towards $\gamma_{PT}$ (center panel) and beyond (bottom panel), the wave packet evolution acquires chirality and the overall intensity also increases from its $\gamma=0$ value. In an open chain, there is only one path from the source to the sink; in contrast, a chain with PBC has two such paths. {\it Physically, the chirality implies that, on average, the path with the higher tunneling amplitude is preferred over the other}. Note that chirality does not represent a preferential flow from source to the sink - the wave packet motion continues past the sink, to the source again - but rather the handedness of the motion. We also emphasize that when $\gamma=0$ (top panel), on average, both paths are equally preferred. These results are robust, independent of the initial location $m_0$ of the wave packet or the distance $(d-1)$ between the impurities. 

The right-hand column in Fig.~\ref{fig:psi} provides a complementary view of chirality development, with corresponding evolution of the reciprocal-space intensity $I_M(u,t)$. The vertical axis in each panel indicates discrete index $1\leq u \leq N$ that translates into $p_u\in(-\pi,\pi]$. The top panel shows that when $\gamma=0$, as the wave packet diffuses, its average momentum is zero. As $\gamma$ approaches $\gamma_{PT}$ (center panel) and beyond (bottom panel), we see that the reciprocal-space intensity develops a pronounced peak at a finite, positive value, consistent with the chirality seen in the left-hand column of the figure. In addition, results at longer times $T\sim 10N\gg N$ show that the reciprocal-space intensity distribution for $\gamma>0$ reaches a steady-state. 

To quantify this chirality and to dissociate it from the exponentially increasing net intensity $I_R(t)=\sum_j I_R(j,t)$ that occurs for $\gamma>\gamma_{PT}$~\cite{kottospower}, we consider the dimensionless, discrete momentum operator on the lattice with PBC,
\begin{equation}
\langle\phi(t)|\hat{p}|\psi(t)\rangle=-\frac{i}{2}\sum_{j=1}^N\frac{(g^*_{j+1}+g^*_j)(f_{j+1}-f_j)}{\sqrt{\langle\phi|\phi\rangle\langle\psi|\psi\rangle}},
\end{equation}
where $|\phi(t)\rangle=\sum_j g_j(t)|j\rangle$, $|\psi(t)\rangle=\sum_j f_j(t)|j\rangle$, and the normalization factor in the denominator is necessary due to the non-unitary time evolution. As is hinted by the right-hand column in Fig.~\ref{fig:psi}, the momentum expectation value $p_\psi(t)=\langle \psi(t)|\hat{p}|\psi(t)\rangle$ in a given state oscillates about zero when $\gamma=0$ and when $\gamma>0$, reaches a steady-state value $p(\gamma)\equiv\int_0^T p(t')dt'/T$. Note that since $|p_\psi(t)|\leq 1$ for any initial state and time, the magnitude of $p(\gamma)$ is bounded by unity. 

\begin{figure}[htb]
\begin{center}
\includegraphics[angle=0,width=\columnwidth]{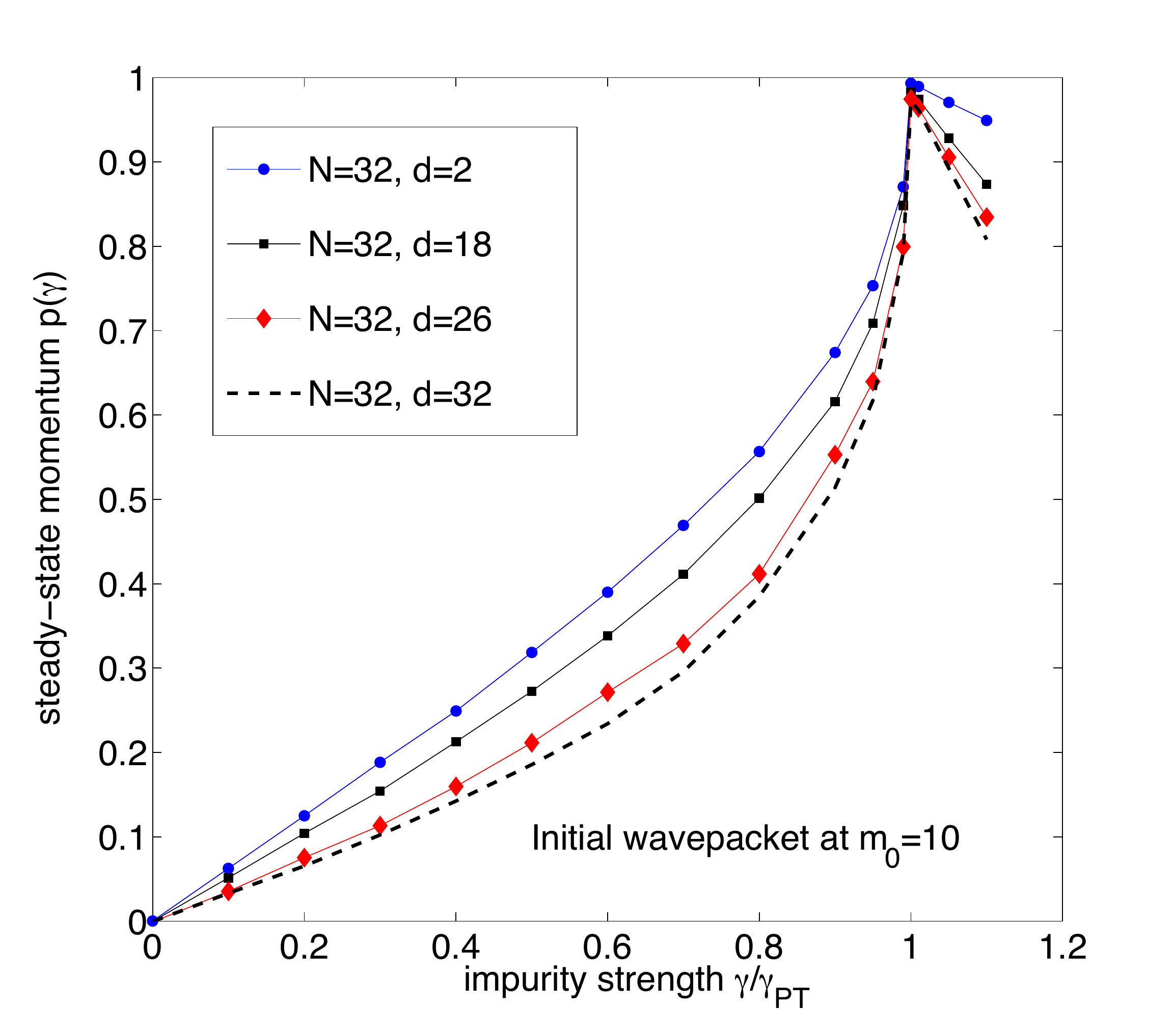}
\caption{(color online) Dependence of chirality, quantified by the steady-state momentum $p(\gamma)$, as a function of impurity strength $\gamma$ for different locations $d$ of the loss impurity $-i\gamma$; the gain impurity $i\gamma$ is located on site 1. The initial wave function is localized on site $m_0=10$. The momentum $p(\gamma)$ varies linearly with $\gamma$ at small $\gamma/\gamma_{PT}\ll 1$. It reaches a universal, maximum value, $p=1$, at the $\mP\mT$-symmetry breaking threshold. For $\gamma/\gamma_{PT}\geq 1$, the steady-state momentum decreases, although the net intensity increases with time. These results are independent of the lattice parameters and the initial wave function.}
\label{fig:pgamma}
\end{center}
\vspace{-5mm}
\end{figure}
Figure ~\ref{fig:pgamma} shows the evolution of the dimensionless, steady-state momentum $p(\gamma)$ across the $\mP\mT$-symmetry breaking threshold for different locations of the loss impurity. These results are for a lattice with $N=32,t_0=0.5,t_b=1$, and thus, $\gamma_{PT}/\max(t_0,t_b)=0.5$. The initial location of the wave packet is $m_0=10$, and we have used normalized time $T=500$ to numerically obtain the average. These results are generic and qualitatively independent of the initial location $m_0$ or the lattice parameters. When $\gamma=0$ there is no chirality to the wave packet evolution and $p(\gamma)=0$. At small $\gamma$, we see that the $p(\gamma)$ increases linearly with $\gamma$, as is expected from a first-order perturbation theory calculation; the slope of this line decreases monotonically with $d$. {\it We find that the steady-state momentum reaches a universal value, $p=1$, at the $\mP\mT$-breaking point and decreases linearly for} $\gamma\geq \gamma_{PT}$. We emphasize that although the total intensity increases exponentially with time, the chirality, which captures the handedness of motion of the wave packet, decreases for  $\gamma>\gamma_{PT}$. 

{\it Discussion.} In this paper, we have investigated the $\mP\mT$ phase diagram and signatures of $\mP\mT$-symmetry breaking in a lattice with periodic boundary conditions. We have presented a model of a lattice with PBC and two uniform tunneling amplitudes, and shown that the $\mP\mT$-symmetric region for such a model is robust, insensitive to the distance between the loss and gain impurities, and widely tunable in size. We have shown that in such a lattice with PBC, where there are two different paths from the source to the sink, the motion of a wave packet acquires a chirality when the impurity strength is nonzero. We have predicted that the $\mP\mT$-symmetry breaking in such a lattice is signaled by a universal, maximal value for the steady-state momentum that quantifies this chirality. 

Traditionally, the investigation of signatures of $\mP\mT$-symmetry breaking has focused on dependence of the intensity profile $I_R(j,t)$ or the net intensity, on time and the impurity strength. These quantities vary smoothly across the phase boundary~\cite{derek,kottospower}. In this paper, we have shown that the chirality, on the other hand, shows a peak with a universal value at the $\mP\mT$-symmetry breaking threshold. We note that such a measurement will require the knowledge of relative phases of the wave function values at adjacent sites, $p(\gamma)\propto\sum_j \Im(f_j^*f_{j+1})$, and therefore is more complex than the site-dependent intensity measurements~\cite{expt1,expt2,christo,berg1}. 

Our results show that $\mP\mT$-symmetric rings~\cite{znojilring} display properties that have no counterparts in open chains. Their in-depth  investigation, including the effects of disorder and non-linearity, will deepen our understanding of $\mP\mT$-symmetric systems with periodic boundary conditions. 


\begin{thebibliography}{99}
\bibitem{caveat} We call a $\mP\mT$-symmetric phase as robust if $\gamma_{PT}(N)$, measured in units of the bandwidth of the $N$-site lattice, is nonzero as $N\rightarrow\infty$.
\end{thebibliography}
\end{document}